# A GPU implementation of a track-repeating algorithm for proton radiotherapy dose calculations

Pablo P Yepes[1], Dragan Mirkovic[2] and Phillip J Taddei[2]

[1] Department of Physics and Astronomy, MS 315, Rice University, 6100 Main Street, Houston, TX 77005, USA
[2] Department of Radiation Physics, Unit 1202, The University of Texas M. D. Anderson Cancer, 1515 Holcombe Blvd., Houston, TX 77030, USA

PACS: 61.80.-x, 61.82.-d, 78.70.-g, 81.40.Wx, 87.53.-j

Key words: proton radiotherapy, track-repeating, GPU, dose calculation, Monte Carlo, fast dose calculator

Corresponding author: Pablo P. Yepes, PhD, Physics and Astronomy Department, MS 315, Rice University, Houston, TX 77005, USA
Phone: 713-248-5942; Fax: 713-348-5215; E-mail: yepes@rice.edu

**Abstract**

An essential component in proton radiotherapy is the algorithm to calculate the radiation dose to be delivered to the patient. The most common dose algorithms are fast but they are approximate analytical approaches. However their level of accuracy is not always satisfactory, especially for heterogeneous anatomic areas, like the thorax. Monte Carlo techniques provide superior accuracy, however, they often require large computation resources, which render them impractical for routine clinical use. Track-repeating algorithms, for example the Fast Dose Calculator, have shown promise for achieving the accuracy of Monte Carlo simulations for proton radiotherapy dose calculations in a fraction of the computation time. We report on the implementation of the Fast Dose Calculator for proton radiotherapy on a card equipped with graphics processor units (GPU) rather than a central processing unit architecture. This implementation reproduces the full Monte Carlo and CPU-based track-repeating dose calculations within 2%, while achieving a statistical uncertainty of 2% in less than one minute utilizing one single GPU card, which should allow real-time accurate dose calculations.

**1. Introduction**

Radiation therapy is an important component of the treatment of cancer. Radiation dose absorbed in normal tissues produces acute effects, for example necrosis, and late effects, such as carcinogenesis. An essential component for the quality of a radiotherapy treatment plan is the accuracy of dose calculations (Papanikolaou et al 2004). The clinical advantages of more accurate dose calculations -tumor recurrence, local control, and normal tissue complications- has not been fully quantified and requires further investigation. Nevertheless evidence exists that dose differences on the order of 7% are clinically detectable (Dutreix 1984). Moreover, several studies have shown that 5% changes in dose can result in 10%-20% changes in tumor control probability or up to 20-30% changes in normal tissue complication probabilities (Orton 1984, Stewart 1975, Goitein 1975).

Dose distributions in proton therapy are typically calculated by commercial treatment planning engines based on analytical pencil-beam algorithms (Petti 1992, Russell 1995, Hong 1996, Deasy 1998, Schneider 1998, Schaffner 1999, Szymanowski and Oelfke 2002.

Schaffner *et al.* (1999) reviewed various analytical proton dose models and concluded that no single pencil-beam model can predict the dose correctly in every situation. They also concluded that the Monte Carlo approach is more accurate than any analytical model and should be used to verify the dose distributions in situations above a certain level of anatomical complexity. Soukup et al (2005) derived a pencil-beam method from Monte Carlo simulations; this method works well for simple heterogeneous or slab-structured phantoms, however it does not achieve the accuracy of the Monte Carlo approach for phantoms describing more complex heterogeneous media, for example head and neck and pelvic geometries. Ciangaru et al (2005) benchmarked analytical calculations of proton doses in simple heterogeneous phantoms and concluded that the algorithms were reasonably accurate for predicting doses at anatomic sites containing laterally extended inhomogeneities that are comparable in density to one another and located away from the Bragg peak. However, the algorithm had mixed success in calculating proton doses in areas with a combination of high and low density media.

The Monte Carlo approach has been shown to provide higher accuracy (Titt et al 2008, Koch et al 2008, Newhauser et al 2008, Paganetti et al 2008, Giebeler et al 2010) than pencil-beam algorithms, however its clinical utilization has been hampered by its high computational requirements. In one study with the Monte Carlo code MCNPX (Pelowitz 2007), for a typical three-beam proton lung cancer, simulating all three proton radiotherapy treatment fields with acceptable statistical precision required up to 5000 hours of central processor unit (CPU) time (Taddei 2009). The high computation times make the use of robust Monte Carlo codes impractical for routine clinical radiotherapy dose calculations, i.e., without the use of large-scale computer clusters.

Various simplified Monte Carlo approaches have also been reported in the literature. Kohno and collaborators (2003) implemented a proton Monte Carlo dose algorithm in which a reduced number of physics processes were taken into account. This approach increased the computational efficiency while largely preserving the accuracy of the calculations (Hotta 2010). Fippel and Soukup (2004) developed another proton Monte Carlo code based on the concept of the voxel Monte Carlo algorithm that had previously been applied to photons and electrons (Fippel 1999). They reported good agreement between dosimetric predictions from their code and from full Monte Carlo codes. Their calculation times were 35 and 13 times faster than Monte Carlo codes GEANT4 (Agostinelli et al 2003; Allison et al 2006) and FLUKA (Fasso et al 2005), respectively. Tourovsky et al (2005) implemented a stochastic proton transport code with simplified physics models and tested it in a variety of clinical cases. Computation times relative to other algorithms were not reported. Li et al (2005) reported a track-repeating algorithm for proton therapy that maintained the accuracy of the Monte Carlo technique, while significantly decreasing computation times. However, their validation was limited to simple heterogeneous phantoms. Extending their work, we developed an alternative track-repeating algorithm, the fast dose calculator (FDC), and applied it to clinical proton treatment planning for the pelvis (Yepes et al. 2009a and Yepes et al. 2009b) and the thorax (Yepes et al. 2010). FDC reproduced the results of GEANT4 simulations to within 2%, yet required less than 1% of the computation time. While the reduction in computation time with FDC was significant, calculation times were still on the order of an hour on a single CPU for the dose calculation of a typical radiotherapy treatment. An alternative, more efficient computation device is needed to use FDC for routine clinical dose verification calculations viable.

Calculations can be accelerated by performing the computation in a parallel manner on multiple processor systems, like CPU clusters or graphics processing units (GPUs). Dose calculations performed on a large number of distributed CPUs has been reported in the literature (Vadapalli 2010). GPU clusters are less expensive and easier to maintain than traditional CPU clusters. GPU-based algorithms have been developed for a variety of tasks in radiotherapy (Jia et al 2010a and 2010b, Gu et al 2009 and 2010, Jacques 2008 and Hissoiny et al 2009). They have been used for dose calculations by Hissoiny et al (2009) and Jacques et al (2008), who implemented superposition convolution algorithms for dose calculation on GPUs, and by Gu et al (2009), who explored the use of GPUs for a finite size pencil beam model. Moreover a Monte Carlo code for coupled electron-photon transport was implemented on a GPU architecture by Jia et al (2010b). They reported speed gains up to a factor of 6.6. However, to date, to our knowledge no attempt to implement a track-repeating algorithm on a GPU architecture has been reported in the literature

In this paper, we report our recent development of the FDC track-repeating algorithm for proton therapy on a GPU architecture under the computer unified device architecture (CUDA) platform developed by NVIDIA (2009). The code has been implemented on a single commercially available graphic card with two GPU units. The GPU-based FDC has been benchmarked against the CPU-based FDC code and the full Monte Carlo GEANT4, since the latter has been validated for proton therapy (Aso et al 2005, Paganetti et al 2008).

## 2. Methods

### 2.1 GEANT4

We used the GEANT4 tool kit (version 4.8.3) (Agostinelli *et al* 2003; Allison *et al* 2006) to generate the database of pre-calculated proton histories and to generate the reference dose sample. The physics models in this setup included proton energy loss *via* the continuous slowing-down approximation for secondary electrons with energy below Tc. Tc is material dependent and calculated from a particle distance range set to 0.1 mm. This translates to 83.6 KeV and 250 eV energy cutoffs for electrons in water and air, respectively. In this work we utilized the low-energy parametrized model (Chauvie 2004), which takes into account atomic and shell effects and is applicable down to 250 eV. It uses the Bethe-Bloch formula to calculate hadron ionization down to 2 MeV, and a ICRU 49 parametrization (ICRU 1993) in the range 1 keV to 2 MeV. Below 1 keV the free electron gas approach is utilized. The energy straggling was calculated with a Gaussian distribution with Bohr's variance (ICRU 1993) for distances long enough for the approximation to be accurate. For short distances a simple model of the atom was used (GEANT4 2008 ).

The multiple Coulomb scattering for protons was estimated with a condensed simulation algorithm, in which the global effects of the collisions are estimated at the end of a track segment. It uses model functions to determine the angular and spatial distributions, which were chosen to reproduce the distributions of the Lewis theory (Lewis 1950*).*

For elastic hadronic interactions the low-energy parametrized model was implemented. Inelastic interactions were simulated with a pre-equilibrium model in the range of interest to our simulations (0-250 MeV). The model is based on Griffin's semi-classical description of composite nucleus decay (Griffin, 1966, Gudima *et al* 1983; Lara and Wellisch 2000). A more detailed description of the models used in proton therapy can be found elsewhere (Jarlskog and Paganetti, 2008).

### 2.2 Fast Dose Calculator

The FDC algorithm utilizes a pre-generated database of the histories of particles produced by a proton impinging on a water phantom. In our study, we have generated the database using the Monte Carlo code GEANT4. Each particle trajectory is broken into steps, and for each step the direction, length and energy loss is stored. FDC calculates dose distributions in heterogeneous anatomies, by re-tracing proton tracks. This re-tracing is achieved by scaling the length and scattering angle of each step according to the material in the non-water medium. In this study the database of proton histories was generated by simulating 121 MeV protons impinging on a 510x510x2500 mm$^3$ water phantom with a 3x3x2500 segmentation along the (x,y,z) axes. Further details of the FDC code were described previously (Yepes 2009a, 2009b).

In this work the dose and deposited energy calculated with GEANT4 and FDC, as described previously (Yepes, 2009a, Yepes 2009b, Yepes 2010), were used as basis of comparison with prediction from GFDC. Results from the CPU version of FDC reported in this work utilized a database with 10 million pre-calculated proton histories in water, while the GPU-based version utilized a database with only 100,000 proton histories. Such a relative small database was chosen because no undesired correlations were observed with such a reduced number of 100,000 proton histories (see Results section). A small database should make the algorithm easier to install on various platforms.

## 2.3 CUDA Implementation

GFDC was developed using the CUDA software platform (NVIDIA 2009) on a general purpose GPU on a single graphics card (GEFORCE GTX 295, NVIDIA, Santa Clara, CA) with 1.79GB of global memory as the hardware platform. That GPU card holds 2 GPU units, with each unit holding 240 GPU cores. The software environment calls for the generation of multiple computational threads. The total number of threads is defined by the programmer and can be as high as hundreds of thousands. Threads are divided into blocks, so that the total number of threads is the number of blocks ($N_B$) multiplied by the number of threads per block ($N_T$). The number of threads should be a multiple of 32 for optimal performance because of hardware configuration of the GPU units. Since the track-repeating algorithm is inherently highly parallel, each thread is treated as an independent computational unit. Each unit re-traces one of the proton histories from the database of pre-calculated histories.

A flowchart of the GPU implementation of FDC is shown in Figure 1. The code was split into sections to be executed on the CPU and on the GPU, with the GPU being called from the CPU code. After the program initialization on the CPU, it first reads the material and geometry information from configuration files. That information was then stored in GPU global memory accessible to the GPU code. The material information corresponds to the scaling parameters of the step length and scattering angle for all the materials to be used in the calculation. The geometry information, derived from the patient CT-scan, consists of a three dimensional array with a material index for each voxel.

After this initial stage, a loop over a pre- determined number of iterations was initiated. For each iteration, a number histories equal to the number of thread blocks ($N_B$) was read from the database. In addition an array with phase-space information for $N_T x N_B$ incident protons to be simulated was generated. This proton phase-space can be read from a file or generated randomly by the program within certain controllable ranges in energy and direction. In the results present in this work, the phase space was generated with fixed energy and direction but a spread in the position transverse to the beam (See section 2.4). Once the phase-space array was generated and the proton histories read, both pieces of information were transferred from the CPU to the GPU global memory, where they could be accessed by all the $N_t x N_B$ GPU threads. Since only $N_B$ pre-calculated proton histories were made available for $N_T x N_B$ threads, the same proton history was utilized $N_T$ times for various positions of the incident protons. However, since trajectories from a given history traversed different areas of the heterogeneous phantoms, the results from the re-tracing of the same history were expected to be statistically independent. Statistical uncertainties are calculated with two alternative methods, as explained in Section 2.3, to test whether different trajectories are statistically independent.

After the operations to initialize an iteration were completed, the GPU code was invoked from the CPU code through a special C-language function termed kernel. The kernel was executed $N=N_T x N_B$ times on the GPU engine in parallel on independent threads. The GPU code was subdivided into two main tasks: 1) finding the database history to be used and selecting the trajectory and step where the track-repeating algorithm should start; 2) re-tracing the pre-calculated proton history through the heterogeneous phantom.

The deposited energy generated by the different threads was tallied in a large three dimensional grid with the same number of voxels as the heterogeneous phantom. The tally grid was defined in GFDC as a one dimensional array and was placed in the GPU global memory. This approach minimized the amount of memory utilized for tallying, which was significant for large grids. However utilizing a common tally for all threads required a function which blocked access to the memory location while a particular thread updated the information stored in it. Blocking a certain memory location forced other threads, trying to update the same voxel tally, to wait until the operation was completed. Such a mechanism slowed down the algorithm when competing threads attempted to access the same memory location simultaneously. A second tallying array, with the same dimension as the tally for the deposited energy, was utilized to store the sum of the squares of the deposited energy. The second array was necessary to estimate the history-by-history statistical uncertainties (See next section). At the end of the run the energy deposited in each voxel was converted to absorbed dose by dividing by the mass of the voxel. The number of blocks ($N_B$) and threads ($N_T$) were optimized to minimize the execution time.

The graphics card used in this study housed two GPU units. In order to maximize the performance of the card, two CPU threads were defined in the code, with each thread handling the operation and data transfer to one of the GPU units. Each CPU thread read from the same database; however, they read different pre-calculated proton histories. At the end of the execution of the two CPU threads, the absorbed dose from both threads was combined.

*2.4. Statistical Uncertainties and Dose Distribution Comparison*

Standard methods (Chetty 2007) were used to calculate statistical uncertainties and to investigate the effects of using the same 100,000 pre-calculated proton histories as many as 250 times. The batch method consists in comparing the results of multiple calculations, or batches, performed with uncorrelated phase space files and random number sequences. For this method, the estimate of uncertainty of the dose, $D$, is given by:

$$\sigma_{\bar{D}} = \sqrt{\frac{\sum_{i=1}^{n}(D_i - \bar{D})^2}{n(n-1)}} \qquad (1)$$

where n is the number of independent batches or runs, $D_i$ is the scored dose in batch $i$, and $\bar{D}$ is the mean value of the absorbed dose over all the batches. The sample size is given by the number of batches or independent calculations. In the history-by-history method, where a history dose corresponds to the absorbed dose produced by a single proton impinging on the phantom, the statistical uncertainty is given by:

$$\sigma_D = \sqrt{\frac{1}{N-1}\left(\frac{\sum_{i=1}^{n} D_i^2}{N} - \left(\frac{\sum_{i=1}^{n} D_i}{N}\right)^2\right)} \qquad (2)$$

where $N$ is number of primary histories, $D_i$ the contribution to the dose by independent history $i$. Whereas the history-by-history approach may be distorted by hidden correlations, the batch method uncertainties would not because each batch is generated with completely independent databases and random numbers. If the estimate of the uncertainties is biased by hidden correlations, we expect a deviation from the $1/\sqrt{N}$ behavior for the history-by-history uncertainties. Moreover the estimated uncertainty for the history-by-history approach will be lower than the unbiased correlations. On the other hand, hidden correlations due a small database are not expected to affect the batch estimates. Thus, comparing the results of these two approaches tests the feasibility of using the same proton history multiple times for these type of calculations.

The mean dose uncertainty was defined as the average of $\sigma_D/D$ over all voxels with a dose larger than half the maximum dose, with $\sigma_D$ calculated with equations (1) and (2).

To quantify the dosimetric accuracy of the FDC and GFDC dose distributions, we compared them to a distribution from GEANT4 simulations for the same field and voxelized phantom. The figure of merit used to quantify the dosimetric accuracy was the gamma index, $\Gamma$ (Low *et al* 1998). This method of evaluating the distance to agreement and dose difference of the sample case versus the reference case is widely used in the comparative analysis of dose distributions in radiotherapy. Two distributions are typically considered to agree well when at least 99% of the voxels, *j*, have values of $\Gamma_j$ smaller than unity. GEANT4, which was previously validated for applications in proton therapy (Aso 2005 et al and Paganetti et al 2008), provided reference dose distribution in this study, against which dose distributions from FDC and GFDC were compared. The maximum acceptable differences in dose and spatial distance used to calculate the $\Gamma$ index in this study were 3% and 3 mm, respectively.

## 2.5. Patient anatomy and radiation field

The geometric model was represented as a voxelized phantom based on the computed tomography (CT) images of the thoracic region of a patient who had been treated for lung cancer at The University of Texas M. D. Anderson Cancer Center. The phantom contained 6,064,305 voxels, each having dimensions of 1x1x2.5 mm$^3$. Each voxel was assigned a material composition and a mass density that corresponded to the Hounsfield unit value in the CT scan for that voxel, following the approach described elsewhere (Newhauser et al 2008). The thoracic region was selected because the thorax is highly inhomogeneous. It is in such in homogeneous areas that pencil-beam algorithms are least reliable.

We have simulated a mono-energetic proton field of 121 MeV and a circular cross section of 4 cm radius. The energy of the beam was selected as to traverse a significant fraction of the lung and, therefore, test the algorithm running on GPUs stringently. The energy and field characteristics were not selected to maximize the dose to the tumor or to minimize the dose to healthy tissue, as this was a generic test field, not a clinically realistic field.

## 3. Results

Figure 2 shows the distribution of deposited energy versus depth (y axis) in the heterogeneous phantom, plotted along the beam central axis, (i.e., x = 0 and z = 0) and 1.5 cm and 3.0 cm lateral to the central beam axis, as predicted by GEANT4, FDC and GFDC. In addition, the percent difference in deposited energy between the track-repeating algorithm (FDC and GFDC) and GEANT4 are plotted along the same axis. Plotting the deposited energy rather than dose was chosen to better show the effects of inhomogeneity of the anatomy. GFDC reproduces the results from the CPU version of the FDC code, the differences can be attributed to statistical fluctuations, since different pre-calculated databases of proton histories were utilized. In addition good agreement was observed between the deposited energy calculated by GEANT4 and GFDC. Figure 3 shows the cross-field profiles in the vertical direction for the isocenter (x = 0 and y = 0) and 7.5 cm posterior and anterior relative to the isocenter (x = 0 and y = +/-7.5 cm). Each profile is calculated along a five voxel thick line. The cross-field profiles are depicted for three penetration depths to illustrate the agreement between the three approaches.

Agreement was excellent for both profiles, with the largest discrepancy less than 3%.

The rest of the results comprise comparisons of dose rather than deposited energy. The difference between the GFDC- and GEANT4-calculated doses for each voxel in the anatomic phantom divided by the maximum GEANT4 voxel dose has a RMS value of 0.5%. From that value we conclude that GFDC reproduces the dosimetric accuracy of GEANT within 1%.

A more comprehensive comparison was obtained by calculating the $\Gamma$-index of FDC and GFDC relative to GEANT4 for each voxel. The $\Gamma$ index results are presented in Figure 4 as the complimentary cumulative distribution function such that the ordinate represents the probability that $\Gamma$ will be greater than the value of the abscissa. Both GFDC and FDC have essentially identical distributions, showing that the GPU-based FDC reproduces the results from the CPU-based version. Moreover, less than 0.01% of the voxels have $\Gamma$ values greater than unity. Thus, the dose distributions from FDC and GFDC are in good agreement with the reference dose distribution from GEANT4.

As explained above, GFDC utilizes a database with 100,000 pre-calculated proton histories and re-traces each history multiple times. In order to verify that such re-cycling of proton histories does not produce undesired statistical correlations, the mean dose uncertainties were calculated with the history-by-history approach and with multiple batches. The batch method utilizes six independent batches or runs. Results were identical if the number of batches was reduced to three. Figure 5 shows the results of the test of re-cycling the proton history database. In the figure, the lines represent a function $f(N) = C/\sqrt{N}$, where C is adjusted for the function to go through the point with the lowest N for each of the two curves. As can be seen in Figure 5, the measured points for both methods follow the $f(N)$ function. If correlations were present, we would expect a deviation from the $1/\sqrt{N}$. Moreover the fact that the batch uncertainties which should not be affected by inter-batch correlations, are smaller than the history-by-history uncertainties demonstrates the absence of undesired statistical correlations introduced by the use of multiple proton histories. As can be seen, uncertainties calculated with the history-by-history method are around 50% higher than those from the batch approach.

The calculation time per proton history was found to depend on the number of blocks ($N_B$) and the number of threads per block ($N_T$). The number of blocks and threads per block were varied within the range allowed by the hardware in order to maximize the algorithm performance. The fastest calculation times were obtained for $N_B$=500 and $N_T$=320, for which 184,525 proton histories were processed per second utilizing the two GPU units on the graphics card. The CPU-based FDC on one CPU processed 2445 proton histories per second. **Therefore the implementation of the FDC algorithm on a GPU card alone achieved a speedup of a factor of 75.5 with respect to the CPU-based implementation.**

Storing the error in the tally arrays was found to increase the calculation time by 18%. In the results reported here errors were not stored. The rationale being that in a clinical environment, error is rarely reported for each voxel.

The upper abscissa of Figure 5 shows the calculation time on the GPU card as a function of the statistical uncertainties. With the batch method a mean statistical dose uncertainty of 1% was achieved in less than one minute, while around 2 minutes were required for the history-by-history approach.

## 4. Discussion

The good agreement between dose distributions calculated with GFDC versus the GEANT4 and FDC codes suggests the feasibility of GFDC to calculate dose distributions in proton radiotherapy as accurately as general purpose Monte Carlo programs. While preserving accuracy, GFDC reduced computational times by a factor of 75 with respect to CPU-based FDC track-repeating algorithm.

Speed gains of 6.6 for a GPU-based Monte Carlo code relative to its CPU-based version were reported in the literature (Jia 2010b). The limited gains of 6.6 for the MC code are thought to be due to the nature of the GPU architecture. On a GPU, the algorithm is executed in multiple threads running in parallel. Threads must run in groups for best performance. Branches in the code do not impact performance provided all threads of a given group follow the same execution path. This may become a significant limitation for any inherently divergent task, like Monte Carlo simulations. It is likely that the speed gains seen for GFDC were large because of the simpler logic of a track-repeating algorithm, as compared to a full Monte Carlo. A simpler logic should generate threads which follow closer execution paths.

In our current GPU algorithm all the threads in a given group were fed with the same proton history from the database of pre-calculated histories. Even though each history is re-traced in different areas of the phantom, and thus produces statistically independent results, this seems to minimize the logic path divergence for the various threads in a group. When threads in a given group were fed with different proton histories, the execution times increased by about 50%.

The CPU-based version used for the results on pelvis anatomy (Yepes 2009a, 2009b) was limited to dose calculations in a voxelized geometry. Results reported for the thorax anatomy included an aperture and range compensator (Yepes 2010). For those results the code included a package from ROOT (Brun 1997) to describe arbitrary geometries. In the GPU-based FDC (GFDC), reported in this study, this feature to describe arbitrary geometries has not been implemented yet, due to the difficulties to port the corresponding ROOT classes to GPUs. The GFDC version is restricted to dose calculation in voxelized geometries.

Currently, the large computational requirement for Monte Carlo simulations makes their use difficult for routine clinical proton radiotherapy treatment planning. At present, it is only practical to calculate such treatment plans with large computer clusters, which are unavailable in most clinics. This obstacle also hinders the opportunities for studies in which whole-body dose reconstructions are needed for large numbers of patients, e.g., in clinical trials or radiation epidemiology studies. The results of the present study suggest that it may be feasible to overcome this obstacle with the GFDC approach, although additional development and testing of the codes will still be needed.

The findings of this study indicate that it may be feasible for GFDC to calculate the dose distribution for an entire proton radiotherapy treatment plan for lung cancer with 1% statistical dosimetric uncertainty on a desktop-size system equipped with 2 GPU cards in about one minute. Future studies to test this hypothesis should include multiple clinically realistic fields, a plurality of patients and sites. The code should also be extended to make it capable to handle arbitrary geometries to include the patient dependent beam shaping elements.

## 5. Conclusions

In conclusion, GFDC is a promising implementation of a track-repeating code for proton radiotherapy dose calculations using a GPU architecture. The dosimetric accuracy of the GFDC algorithm was validated by comparing the results with those generated with GEANT4 Monte Carlo and CPU-based FDC simulations. GFDC can calculate the dose distribution produced by a 120 MeV proton beam in a thoracic geometry with 1% accuracy in one 1 minute with a graphics card with two GPU units. Only 0.01% of the phantom voxels had a $\Gamma$ index larger than one, with the $\Gamma$ index calculated with a full Monte Carlo as the reference distribution. The implementation of the track-repeating algorithm on a GPU architecture may allow for real-time dose calculations for proton radiotherapy with a desktop-size computer system equipped with multiple GPU cards.

**Acknowledgements**

This work was supported in part by the John & Ann Doerr Fund for Computational Biomedicine under contract CABC-2006-5-PY; by the Rice Computational Research Cluster funded by the National Science Foundation under Grant CNS-0421109 and a partnership between Rice University, Advanced Micro Devices., and Cray Inc. This work was supported in part by the National Cancer Institute (award 1R01CA131463-01A1) and by the Fogarty International Center (award K01TW008409). We are thankful to Wayne Newhauser for his careful reading of the manuscript. We are indebted to Vivek Sarkar and Yonghong Yan for discussions on the technical aspects of the GPU implementation. The content is solely the responsibility of the authors and does not necessarily represent the official views of the sponsors.

# References

Agostinelli S *et al* 2003 GEANT4–a simulation toolkit *Nucl. Instrum. Meth. A* **506** 250–303

Allison J *et al* 2006 GEANT4 developments and applications *IEEE Trans. Nucl. Sci.* **53** 270-8

Aso T, Kimura A, Tanaka S, Yoshida H, Kanematsu N, Sasaki T, and Akagi T 2005 Verification of the dose distributions with GEANT4 simulation for proton therapy *IEEE Trans. on Nucl. Sci.* **52** 896-901.

Brun R and Rademakers F 1997 ROOT-an object oriented data analysis framework *Nucl. Instrum. Meth. A* **389** 81-86

Chauvie S, Guatelli S, Ivanchenko V, Longo F, Mantero A, Mascialino B, Nieminen P, Pandola L, Parlati S, Peralta L, Pia M G, Piergentili M, Rodrigues P, Saliceti S, Tnndade A 2004 Geant4 low energy electromagnetic physics *IEEE Nuclear Science Symp. Conf. Rec.* **3** 1881-5

Chetty J T, Curran B, Cygler J E, DeMarco J J, Ezzell G, Faddegon B A, Kawrakow I, Keall PJ, Liu H, Ma CM, Rogers DW, Seuntjens J, Sheikh-Bagheri D and Siebers JV 2007 Report of the AAPM Task Group No. 105: Issues associated with clinical implementation of Monte Carlo-based photon and electron external beam treatment planning *Med Phys* **34**, 4818-53.

Ciangaru G, Polf J C, Bues M and Smith A R 2005 Benchmarking analytical calculations of proton doses in heterogeneous matter *Med. Phys.* **32** 3511-23

Deasy J O 1998 A proton dose calculation algorithm for conformal therapy simulations based on Moliere theory of lateral deflections *Med. Phys.* **25** 476–83

Dutreix A 1984 When and how can we improve precision in radiotherapy? *Radiother. Oncol.* **2** 275-92.

Fasso A, Ferrari A, Ranft J and Sala P R 2005 FLUKA: a multi-particle transport code *CERN-2005-10, INFN/TC_05/11*

Fippel M 1999 Fast Monte Carlo dose calculation for photon beams based on the VMC electron algorithm *Med. Phys.* **26** 1466-75

Fippel M and Soukup M 2004 A Monte Carlo dose calculation algorithm for proton therapy *Med. Phys.* **31** 2263-73

GEANT4 2008 Physics Reference Manual, Version: geant4 9.2

Giebeler A, Zhu X R, Titt U, Lee A, Tucker S and Newhauser 2010 Uncertainty in dose per monitor unit estimates for passively scattered proton therapy, Part I: The role of FCSPS in the prostate, submitted to *Phys. Med. Bio*.

Goitein M and Busse J 1975 Immobilization error: Some theoretical considerations *Radiology* **117** 407-12

Griffin J J 1966 Semiclassical model of intermediate structure *Phys. Rev. Lett.* **17** 478-81

Gu X, Choi D, Men C, Pan H, Majumdar A and Jiang S B 2009 GPU-based ultra fast dose calculation using a finite size pencil beam model *Phys. Med. Biol.* **54** 6287–97

Gu X, Pan H, Liang Y, RCastillo R, Yang D, Choi D, Castillo E, Majumdar A, Guerrero T and Jiang S B 2010 Implementation and evaluation of various demons deformable image registration algorithms on a GPU *Phys. Med. Biol.* **55** 207–19

Gudima K K, Mashnik S G and Toneev V D 1983 Cascade-Exciton Model of Nuclear-Reactions *Nuclear Physics A* **401** 329-61

Hissoiny S, Ozell B and Despres P 2009 Fast convolution-superposition dose calculation on graphics hardware *Med. Phys.* **36** 1998–2005

Hong L, Goitein M, Bucciolini M, Comiskey R, Gottschalk B, Rosenthal S, Serago C and Urie M 1996 A pencil beam algorithm for proton dose calculations *Phys. Med. Biol.* **41** 1305–30

Hotta K, Kohno R,Takada Y, Hara Y, Tansho R, Himukai T, Kameoka S, Matsuura T , Nishio T and Ogino T 2010 Improved dose-calculation accuracy in proton treatment planning using a simplified Monte Carlo method verified with three-dimensional measurements in an anthropomorphic phantom *Phys. Med. Biol.* **55** 3545–56

ICRU (International Commission on Radiation Units and Measurements) 1993 Stopping Powers and Ranges for Protons and Alpha Particles *Report 49.*

Jacques R, Taylor R, Wong J and McNutt T 2008 Towards real-time radiation therapy: GPU accelerated superposition/convolution High-Performance Medical Image Computing and Computer Aided Intervention Workshop, HP-MICCAI 2008

Jarlskog CZ and Paganetti H 2008 Physics Settings for Using the Geant4 Toolkit in Proton Therapy *IEEE Trans. Nucl. Sci.* **55** 1018-25

Jia X, Lou Y, Li R, Song WY and Jiang S B 2010a GPU-based fast cone beam CT reconstruction from undersampled and noisy projection data via total variation *Med. Phys.* **37** 1757–60

Jia X, Gu X , Sempau J , Choi D, Majumdar A  and Jiang S B 2010b Development of a GPU-based Monte Carlo dose calculation code for coupled electron–photon transport *Phys. Med. Biol.* **55** 3077–86

Koch N, Newhauser WD, Titt U, Gombos D, Coombes K, Starkschall G 2008 Monte Carlo calculations and measurements of absorbed dose per monitor unit for the treatment of uveal melanoma with proton therapy *Phys. Med. Biol.* **53** 1581-94

Kohno R, Takada Y, Sakae T, Terunuma T, Matsumoto K, Nohtomi A and Matsuda H 2003 Experimental evaluation of validity of simplified Monte Carlo method in proton dose calculations *Phys. Med. Biol.* **48** 1277–88

Lara V and Wellisch JP 2000 Preequilibrium and equilibrium decays in Geant4 in Proc. Computing in High Energy and Nuclear Physics, Padova, Italy, 52-55.

Lewis. H W 1950 Multiple Scattering in an Infinite Medium *Phys. Rev.* **78** 526

Li J S, Shahine B, Fourkal E and Ma C M 2005 A particle track-repeating algorithm for proton beam dose calculation *Phys. Med. Biol.* **50** 1001–10

Low D A, Harms W B, Mutic S and Purdy J A 1998 A technique for the quantitative evaluation

of dose distributions *Med. Phys.* **25** 656–61

Newhauser WD, Zheng Y, Taddei PJ, Mirkovic D, Fontenot J, Giebeler A, Zhang R, Titt U and Mohan R 2008 Monte Carlo proton radiation therapy planning calculations *Trans. Am. Nucl. Soc.* **99** 63-4

NVIDIA 2009 NVIDIA CUDA Compute Unified Device Architecture, Programming Guide

Orton C G, Mondalek P M, Spicka J T, Herron D S, and Andres L I 1984 Lung corrections in photon beam treatment planning: Are we ready? *Int. J. Radiat. Oncol. Biol. Phys.* **10** 2191-99.

Paganetti H, Jiang H, Parodi K, Slopsema R and Engelsman M 2008 Clinical implementation of full Monte Carlo dose calculation in proton beam therapy *Phys. Med. Biol.* **53** 4825-53

Papanikolaou N, Battista J, Boyer A, Kappas C, Klein E, Mackie T, Sharpe M and Van Dyk, J 2004 Tissue inhomogeneity corrections for megavoltage photon beams, *AAPM Report* No. *85*.

Pelowitz DB, ed. 2007 MCNPX™ User's Manual, Version 2.6.0. Los Alamos, New Mexico: Los Alamos National Laboratory

Petti P L 1992 Differential-pencil-beam dose calculations for charged particles *Med. Phys.* **19** 137–49

Russell K R, Grusell E and Montelius A 1995 Dose calculations in proton beams: range straggling corrections and energy scaling *Phys. Med. Biol.* **40** 1031–43

Schaffner B, Pedroni E and Lomax A 1999 Dose calculation models for proton treatment planning using a dynamic beam delivery system: an attempt to include density heterogeneity effects in the analytical dose calculation *Phys. Med. Biol*. **44** 27–41

Schneider U, Schaffner B, Lomax A J, Pedroni E and Tourovsky A 1998 A technique for calculating range spectra of charged particle beams distal to thick inhomogeneities *Med. Phys.* 25 457–63

Soukup M, Fippel M and Alber M 2005 A pencil beam algorithm for intensity modulated proton therapy derived from Monte Carlo simulations *Phys. Med. Biol.* **50** 5089–104

Stewart J G and Jackson A W 1975 The steepness of the dose response curve both for tumor cure and normal tissue injury *Laryngoscope* **85** 1107-11

Szymanowski H and Oelfke U 2002 Two-dimensional pencil-beam scaling: an improved proton dose algorithm for heterogeneous media *Phys. Med. Biol.* **47** 3313–30

Taddei PJ, Mirkovic D, Fontenot JD, Giebeler A, Zheng Y, Kornguth D, Mohan R,

Newhauser WD 2009 Stray radiation dose and second cancer risk for a pediatric patient receiving craniospinal irradiation with proton beams *Phys. Med. Biol.* **54** 2259-75

Titt U, Sahoo N, Ding X, Zheng Y, Newhauser WD, Zhu XR, Polf JC, Gillin MT, Mohan R. 2008 Assessment of the accuracy of an MCNPX-based Monte Carlo simulation model for predicting three-dimensional absorbed dose distributions *Phys. Med. Biol.* **53** 4455-70

Tourovsky A, Lomax AJ, Schneider U and Pedroni E 2005 Monte Carlo dose calculations for spot scanned proton therapy *Phys. Med. Biol*. **50** 971–81

Vadapalli R, Yepes P, Newhauser W and Licht R 2010 Grid-Enabled Treatment Planning for Proton Therapy Using Monte Carlo Simulations *Nucl. Technol.* in press.

Yepes P, Randeniya S, Taddei P and Newhauser W 2009a A track repeating algorithm for fast Monte Carlo dose calculations of proton radiotherapy *Nucl. Technol.* **168** 334-7

Yepes P, Randeniya S, Taddei P and Newhauser W 2009b Monte Carlo fast dose calculator for proton radiotherapy: application to a voxelized geometry representing a patient with prostate cancer *Phys. Med. Biol.* **54** N21-N28

Yepes P, Brannan T, Huang J, Mirkovic D, Newhauser W D, Taddei P J, Titt U 2010 Application of a fast proton dosecalculation algorithm to a thorax geometry *Rad. Meas.* In press.

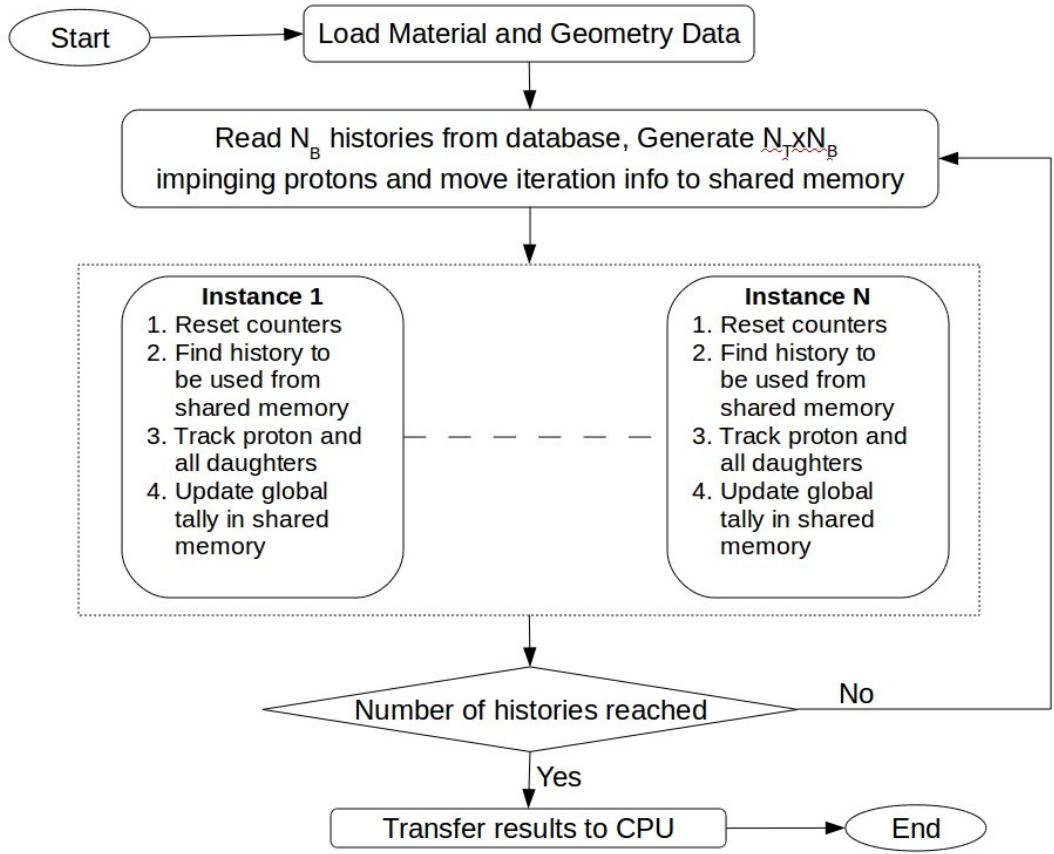

*Figure 1:* The flow chart of the GPU-based version of the track-repeating Fast Dose Calculator (GFDC). The kernel, code running on the GPUs, is bounded by the pointed-line rectangle, and is run in parallel by $N=N_T \times N_B$ different CUDA threads, where $N_B$ and $N_T$ are the number of blocks and threads per block, respectively.

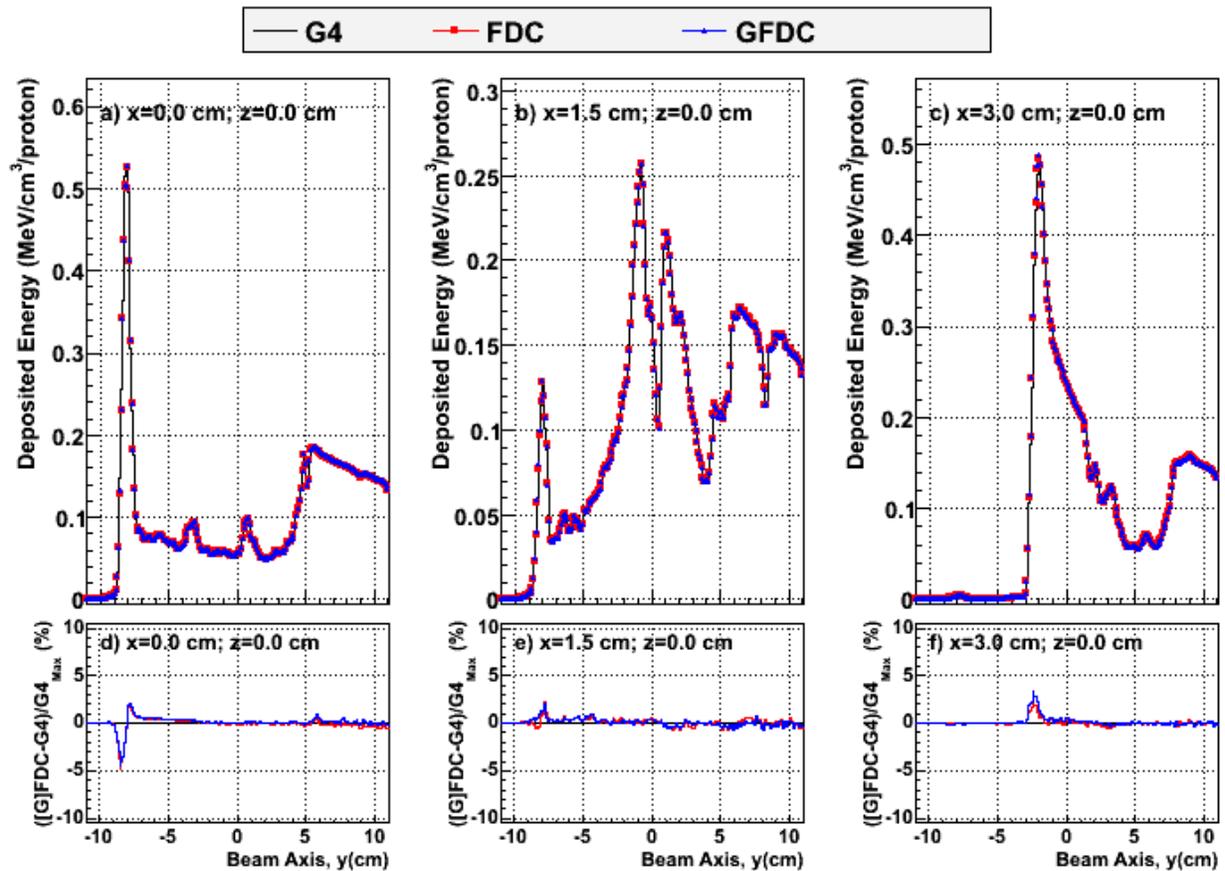

*Figure 2:* Deposited energy profiles in voxels along the beam axis, y, for z = 0 and (a) x = 0, (b) x=1.5 cm and (c) x = 3.0 cm. The y axis runs from posterior to anterior of the patient. Distributions were calculated with GEANT4 (G4: black line), FDC (red circles), and GFDC (blue triangles). The differences in dose between GEANT4 and FDC (red line) and GFDC (blue line) and GEANT4 divided by the maximum GEANT4 dose are shown in panels (d), (e) and (f) for x = 0, x = 1.5 cm, and x = 3.0 cm, respectively.

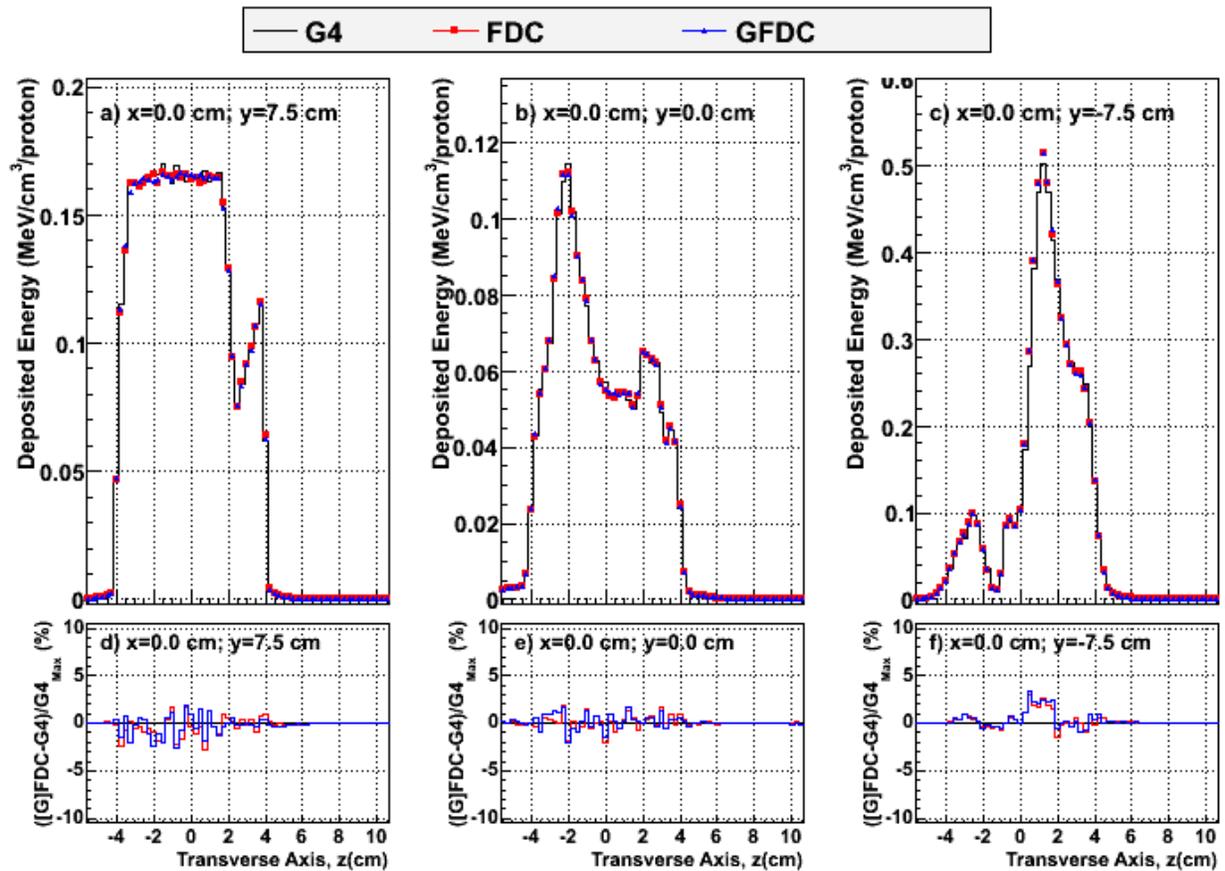

*Figure 3:* Cross-field profiles of deposited energy along the patient's vertical axis (z), along the central beam axis (i.e., x = 0) for (a) y = 7.5 cm, (b) y = 0 cm and (c) y= -7.5 cm, where the y-axis runs from anterior to posterior of the patent. Distributions were calculated with GEANT4 (G4: black line), FDC (red circles) and GFDC (blue triangles). The GEANT4-FDC (red line) and GEANT4-GFDC (blue line) deposited energy differences divided by the maximum GEANT4 deposited energy is shown in panels (d), (e) and (f) for y= 7.5 cm, y=0 cm, and y = -7.5 cm, respectively.

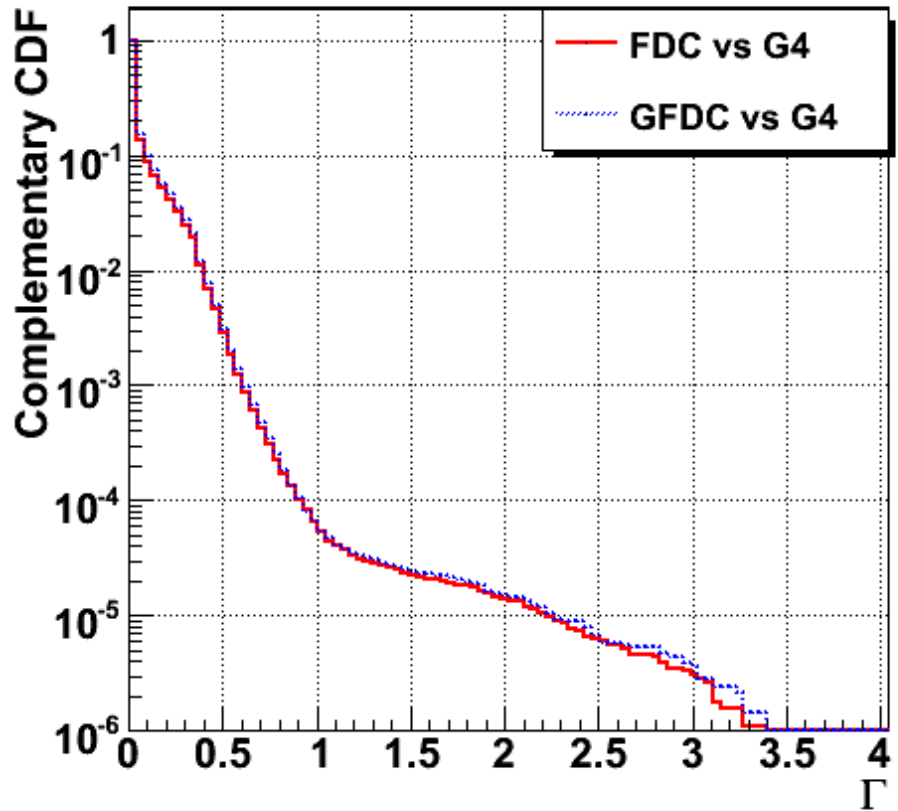

*Figure 4:* The complementary cumulative distribution function (CDF) of the Γ index for FDC and GFDC using GEANT4 as the best estimate of the true dose distribution in the heterogeneous phantom representing a thoracic cancer patient. The gamma function was calculated for all non-air voxels in the geometric model.

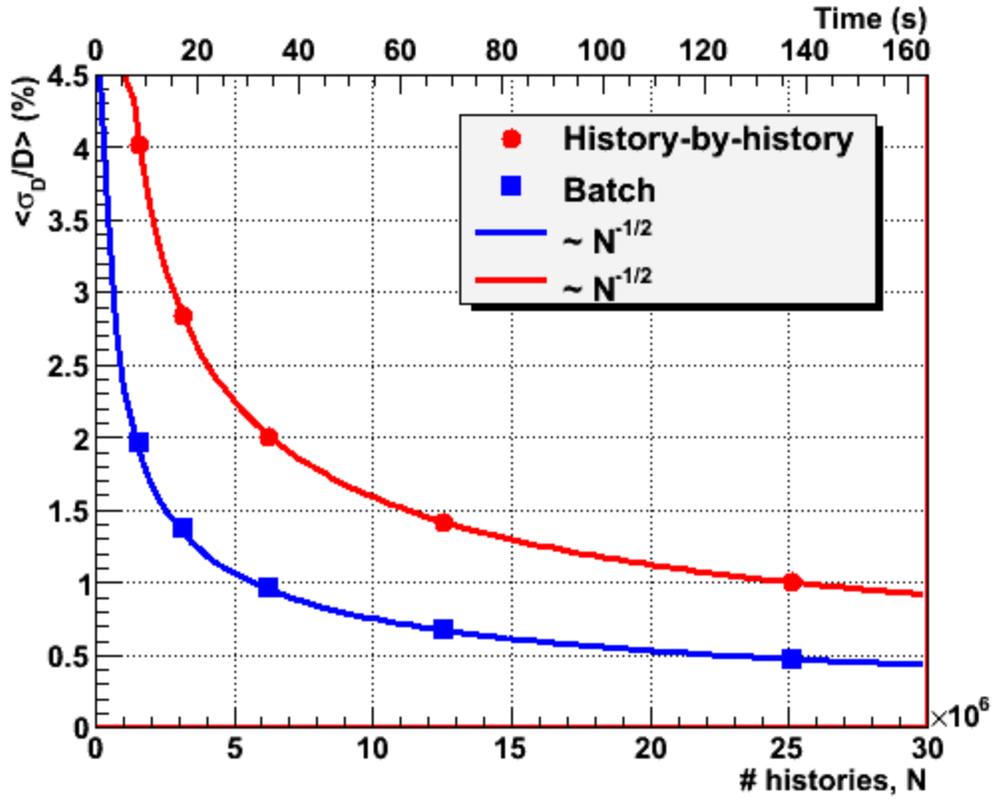

*Figure 5:* The mean statistical uncertainties of the GFDC dose distributions calculated with the history-by-history and the batch approaches as a function of the number of proton histories (N) and calculation times. The lines are functions proportional to $N^{-1/2}$ adjusted to cross the point with the lowest N for each of the methods.